\documentclass[twocolumn,showpacs,pra,aps,superscriptaddress,floatfix]{revtex4-2}%
\usepackage{amsmath}
\usepackage[dvips]{graphicx}
\usepackage{upgreek}
\usepackage{dcolumn}
\usepackage{makeidx}
\usepackage{color}
\usepackage{float}
\usepackage{mathtools}
\usepackage{threeparttable}
\usepackage[linkcolor=blue,anchorcolor=black,citecolor=blue]{hyperref}%

\newcommand{\ket}[1]{\left|#1\right>}
\newcommand{\bra}[1]{\left<#1\right|}
\newcommand{\bk}[2]{\left<#1|#2\right>}
\renewcommand{\L}[1]{\mathcal{L}\left\{#1\right\}(s)}

\newcommand{\pp}{\rho_A}
\newcommand{\pq}{\rho_{\bar{A}}}

\begin{document}
\author{A.~Ricottone}
\email[]{alessandro.ricottone@mail.mcgill.ca}
\affiliation{Department of Physics, McGill University, Montr\'eal, Qu\'ebec H3A 2T8, Canada}

\author{M.~S.~Rudner}
\email{rudner@nbi.ku.dk}
\affiliation{Niels Bohr International Academy and Center for Quantum Devices, University of Copenhagen, 2100 Copenhagen, Denmark}

\author{W.~A.~Coish}
\email{coish@physics.mcgill.ca}
\affiliation{Department of Physics, McGill University, Montr\'eal, Qu\'ebec H3A 2T8, Canada}

\title{Topological transition of a non-Markovian dissipative quantum walk}

\date{\today}

\begin{abstract}
We extend non-Hermitian topological quantum walks on a Su-Schrieffer-Heeger (SSH) lattice [M.~S.~Rudner and L.~Levitov, Phys.~Rev.~Lett.~\textbf{102}, 065703 (2009)] to the case of non-Markovian evolution.  This non-Markovian model is established by coupling each unit cell in the SSH lattice to a reservoir formed by a quasi-continuum of levels.  We find a topological transition in this model even in the case of non-Markovian evolution, where the walker may visit the reservoir and return to the SSH lattice at a later time.  The existence of a topological transition does, however, depend on the low-frequency properties of the reservoir, characterized by a spectral density $J(\epsilon)\propto |\epsilon|^\alpha$. In particular, we find a robust topological transition for a sub-Ohmic ($\alpha<1$) and Ohmic ($\alpha=1$) reservoir, but no topological transition for a super-Ohmic ($\alpha>1$) reservoir.  This behavior is directly related to the well-known localization transition for the spin-boson model. We confirm the presence of non-Markovian dynamics by explicitly evaluating a measure of Markovianity for this model.
\end{abstract}
\maketitle
\section{Introduction}

Quantum walks represent the quantum version of classical random walks \cite{aharonov1993quantum,kempe2003quantum,venegas2012quantum}, where a particle, the ``walker,'' moves on a lattice evolving into a superposition of states. The walker may possess internal degrees of freedom in addition to the spatial degree of freedom.  Translation of the particle may then depend on its internal state. Quantum walks can be used as a platform for universal quantum computation \cite{childs2009universal}, or to study topological phases \cite{kitagawa2010exploring, ramasesh2017direct, flurin2017observing}. For quantum walks, a topological phase can be inferred from the observation of boundary states \cite{kitagawa2012observation,meier2016observation}, from bulk observables such as the mean chiral displacement \cite{cardano2017detection,maffei2018topological}, or directly from a winding number \cite{rudner2009topological}.
Quantum walks have been extended using the language of open quantum systems \cite{whitfield2010quantum,attal2012open}, with applications to photosynthetic energy transfer \cite{mohseni2008environment,rebentrost2009environment}, and environment-assisted transport \cite{maier2019environment}. Quantum walks have also been extended to account for non-Markovian evolution \cite{benedetti2016non, kumar2018non}, describing information back-flow from an environment. 

In non-Hermitian quantum walks, the walker can escape from the lattice with a decay rate showing up as a non-Hermitian term in the Hamiltonian \cite{rudner2009topological,rakovszky2017detecting,lieu2018topological}.  Early work on non-Hermitian quantum walks extended the Su-Schrieffer-Heeger (SSH) model in this way by including an on-site imaginary energy, and introduced  a bias potential on sublattice sites. The SSH model consists of a one-dimensional chain with dimerized (alternating) tunnel coupling between sites, one for intracell hopping and another for intercell hopping.  In the non-Hermitian topological quantum walk based on the extended SSH lattice \cite{rudner2009topological}, the average displacement achieved by the walker before leaving the lattice is a topological invariant, given by a winding number. This winding number changes only at a critical ratio of the intercell to intracell coupling and is otherwise unaffected by details of the system \cite{rudner2009topological,zeuner2015observation}. In $k$-space, the tunnel coupling becomes a complex number described by a magnitude and a phase. The winding number represents the number of times this phase winds as $k$ traverses the Brillouin zone.
This winding number is important, in general, to classify the topological phases of non-Hermitian one-dimensional systems \cite{rudner2016survival}.

In this paper we derive the condition for a topological transition in an alternative extended version of the SSH model where, in each unit cell, a lattice site is coupled to a reservoir formed by a quasi-continuum of levels.  We study the resulting dynamics of the walker in this extended model without performing the Markov approximation. Recently, other extensions of the SSH model that include additional sites in each unit cell have been studied, and the chiral displacement has been analyzed in these models \cite{maffei2018topological}.  
Here, we go beyond this analysis allowing for a general structure of the reservoir having a quasi-continuous bath spectral density, which reduces to the model of Ref.~\onlinecite{rudner2009topological} in a well-defined Markovian limit.
In fact, we find (for this generalized model) that the presence of a topological transition is determined through the low-frequency properties of the reservoir, characterized by the spectral density $J(\epsilon)\propto |\epsilon|^\alpha$, with exponent $\alpha$. In particular, we find a robust topological transition for a sub-Ohmic ($\alpha<1$) and Ohmic ($\alpha=1$) reservoir, but no topological transition for a super-Ohmic ($\alpha>1$) reservoir.  This behavior is directly related to the well-known localization transition in the spin-boson model \cite{leggett1987dynamics}. 

Remarkably, even in the case of non-Markovian evolution, where the walker may visit the reservoir and return, we find a robust topological phase transition in terms of the time-averaged mean displacement, which again can be written in terms of a winding number.  We confirm the presence of non-Markovian dynamics in this model directly through Breuer's measure of Markovianity \cite{breuer2009measure}. This problem is analogous to the Weisskopf-Wigner model for the spontaneous decay of a two-level atom coupled with a continuum of radiation modes \cite{scully1999quantum}. Thus, it may be possible to probe these results in experiments on stimulated Raman adiabatic passage (STIRAP) \cite{vitanov2017stimulated}, or in cold-atom systems with a controllable spectral density \cite{krinner2018spontaneous}.

This paper is organized as follows: In Sec.~\ref{sec:ssh} we present the lattice and the Hamiltonian, and derive the conditions under which the topological invariant of interest, the long-time time-averaged displacement, can be obtained in terms of the winding number. In Sec.~\ref{sec:equivalent_description}, the problem is mapped to the Weisskopf-Wigner model, as a special case of the spin-boson model. This will prove to be useful once the low-frequency properties of the environment are described in terms of a spectral density $J(\epsilon)\propto |\epsilon|^\alpha$, a problem already encountered for the spin-boson model \cite{leggett1987dynamics}. In Sec.~\ref{sec:ltd} we lay down the quantum-dynamical formalism used to calculate the topological invariant found in Sec.~\ref{sec:ssh}. In Sec.~\ref{sec:nMk} we study the non-Markovian properties of the system using a measure derived in Ref.~\onlinecite{breuer2009measure}. In Sec.~\ref{sec:con} we draw conclusions and discuss possible experimental realizations of the model.

 \section{Quantum walk Hamiltonian}
 \label{sec:ssh}
 
\begin{figure}
\centering
\includegraphics[width=\columnwidth]{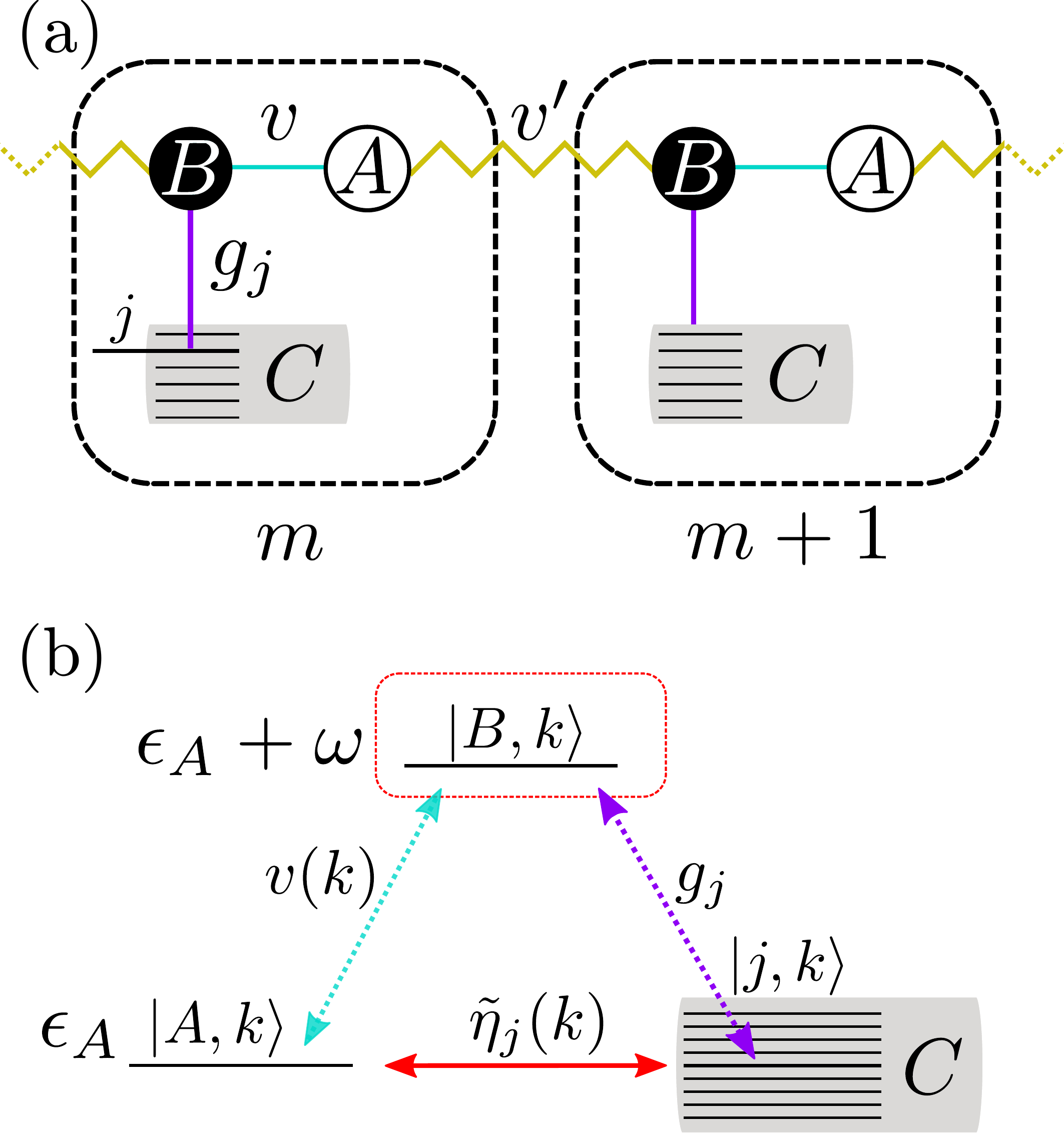}
\caption{(a) The lattice defining the quantum walk. Each unit cell contains 3 sites, $A$ (open circles), $B$ (filled circles), and a quasi-continuum of levels $j$ labeled collectively as $C$ (gray box). Sites $A$ and $B$ are coupled via real-valued intracell ($v$) and intercell ($v'$) tunneling amplitudes. The site $B$ is also coupled to each level in $C$, with coupling $g_j$. (b) Level structure of the system in $k$-space after performing a Schrieffer-Wolff transformation that eliminates the level $\ket{B,k}$ for large detuning $\omega$. The level $\ket{A,k}$ couples to the levels $\ket{j,k}$ in the reservoir with renormalized couplings $\tilde{\eta}_j(k)$. }
\label{fig:lattice} 
\end{figure}

The quantum walk that we study describes hopping of a quantum particle on a translationally invariant one-dimensional chain with unit cells labeled by an integer $m$ and a staggered (dimerized) coupling between $A$ and $B$ sub-lattice sites 
[see Fig.~\ref{fig:lattice}(a)].
The dense set of levels $C$ explicitly models a reservoir that the walker can enter and exit via state $B$ as it moves through the lattice. 
We therefore label the state of the walker by $m$ and $l\in \left\{A,B,j\right\}$, where $j$ indexes the levels within $C$.
This model generalizes the non-Hermitian quantum walk studied in Ref.~\onlinecite{rudner2009topological}, where the walker could only leave the lattice from sites $B$ without the possibility to return. 

In direct analogy with the non-Hermitian quantum walk, for the present model we take the escape probability to be the probability to leave sublattice $A$, i.e., the probability to be either on a $B$ site or in one of the levels $C$.  In the rest of this paper, we will group $B$ sites and $C$ levels in each unit cell into a common `reservoir.'
The observable of interest, the average displacement, is the average distance (number of unit cells) traveled by the walker before leaving sublattice $A$ (reaching a reservoir, i.e., sublattice sites $B$ or $C$).   This leads to the central question of this work: Is the topological transition found in Ref.~\onlinecite{rudner2009topological} still present when we allow for non-Markovian dynamics of the walker? 

 We express the quantum walk Hamiltonian as a sum of terms, $H =  H_{A}+H_{\bar{A}} + V$, with
\begin{align}
H_{A} &= \sum\limits_{m=-\infty}^\infty\epsilon_A \ket{A,m}\bra{A,m},\label{eq:HA}\\
H_{\bar{A}} &= \sum\limits_{m=-\infty}^\infty(\epsilon_A+\omega) \ket{B,m}\bra{B,m} +  \sum_{j\in C}\epsilon_j\ket{j,m}\bra{j,m}\nonumber\\
&+  \sum_{j\in C}  \left(g_j\ket{j,m}\bra{B,m} + \mathrm{h.c.}\right),\label{eq:HQ}\\
V &=\sum\limits_{m=-\infty}^\infty v \ket{A,m}\bra{B,m}+v' \ket{A,m}\bra{B,m+1}+\mathrm{h.c.}.
\end{align}
Here, $H_{A}$ gives the energy of state $A$,
$H_{\bar{A}}$ describes the reservoir (i.e., state $B$ and states $C$), and $V$ contains the amplitudes $v$ and $v'$ for intracell and intercell hopping between states $A$ and $B$, respectively. The amplitudes $v,v'$ are both taken to be real. 
We take the energy of state $B$ to be detuned with respect to the energy of state $A$ by $\omega$. This setting is analogous to the STIRAP problem \cite{vitanov2017stimulated}, where one aims to coherently transfer population from an initial ground state (here $A$) to to a final state (a specific linear combination of states $\left\{j\right\}$ in $C$), exploiting the coupling to an excited state (here $B$) without ever occupying it.

Due to the translational invariance of the quantum walk, in our analysis below it will be convenient to work in momentum space. To this end we introduce the Fourier transforms
$\ket{l,k}=\frac{1}{\sqrt{2\pi}}\sum_{m=-\infty}^\infty e^{ikm}\ket{l,m}$ 
 and $\ket{l,m}=\oint \frac{dk}{\sqrt{2\pi}}\; e^{-ikm}\ket{l,k}$, 
where $\oint dk \rightarrow \int_{-\pi}^\pi dk$ denotes integration over the Brillouin zone and $l$ ranges over $A$, $B$ and $j\in C$.
The Hamiltonian can thus be expressed as $H = \oint dk\,[H_A(k)+H_{\bar{A}}(k) + V(k)]$, with
 \begin{align}
H_A(k) &= \epsilon_A \ket{A,k}\bra{A,k},\label{eq:HAk}\\
H_{\bar{A}}(k) &= (\epsilon_A+\omega) \ket{B,k}\bra{B,k} +  \sum_{j\in C}\epsilon_j\ket{j,k}\bra{j,k}\nonumber\\
&+  \sum_{j\in C}  \left(g_j\ket{j,k}\bra{B,k} + \mathrm{h.c.}\right),\label{eq:HQk}\\
\label{eq:Vk} V(k) &=|v(k)| \left(e^{i\phi(k)}\ket{A,k}\bra{B,k} + \mathrm{h.c.}\right).
\end{align}
Here, the angle $\phi(k)$ is the argument of the complex hopping amplitude $v(k) = v + v'e^{ik}$: 
 \begin{align}
 \phi(k)&=\mathrm{arg}\left[v(k)\right]=\mathrm{arg}\left[v\left(1+ue^{ik}\right)\right],\quad u=\frac{v'}{v}.
 \end{align}
The parameter $u$, the ratio of intracell and intercell hopping amplitudes, controls the topological transition that was investigated in the related non-Hermitian quantum walk studied in Ref.~\onlinecite{rudner2009topological}.

\section{Quantum walk evolution}
We now analyze the evolution of a quantum walker initialized in state $A$ within unit cell $m = 0$, 
\begin{align}
\psi_{l m}(0)&= \bk{l, m}{\psi(0)}=\delta_{l,A}\delta_{m,0}.\label{eq:init_m}
\end{align}
Our goal is to evaluate the average number of unit cells traversed by the walker before leaving sublattice $A$ in the long-time limit, under evolution governed by Hamiltonian $H$ defined in Sec.~\ref{sec:ssh}.  To evaluate this quantity, we first define a set of probabilities 
$\{\rho_{\bar{A}m}(t)\}$, where 
\begin{align}
\label{eq:probs}\rho_{\bar{A}m}(t) &= \sum\limits_{l\notin A}\left|\psi_{l m}(t)\right|^2
\end{align}
gives the probability 
that the walker resides in the reservoir of unit cell $m$ at time $t$.

In terms of the probabilities in Eq.~(\ref{eq:probs}), the average displacement at time $t$ is given by
\begin{equation}
\left<m(t)\right>=\sum\limits_m m \rho_{\bar{A}m}(t).\label{eq:m_m}
\end{equation}
In $k$-space, Eq.~\eqref{eq:m_m} can be rewritten using the probability density $\rho_{\bar{A}}(k, t)$ at quasi-momentum $k$ at time $t$ (see Appendix \ref{app:avgD}):
\begin{align}
\left<m(t)\right>&=\oint dk \; \frac{\partial\phi(k)}{\partial k}\pq (k,t),\label{eq:mk}
\end{align}
where
\begin{align}
\pq(k, t)&=\sum\limits_{l\notin A}\left|\bk{l,k}{\psi(t)}\right|^2\equiv\sum\limits_{l\notin A}\left|\psi_{l}(k,t)\right|^2.\label{eq:escapeProb} 
\end{align}

Non-Markovian (history-dependent) dynamics can give rise to non-decaying persistent oscillations of the probability density $\pq(k, t)$ [see, e.g., Fig.~\ref{fig:probability_time}(b) below]. Hence, in general, the long-time limit of Eq.~\eqref{eq:mk} is not necessarily well defined. As an alternative, we characterize the average displacement before leaving sublattice $A$ by the long-time time-averaged value $\left<m\right>$:
\begin{align}
\left<m\right> &= \lim\limits_{T\rightarrow\infty}\frac{1}{T}\int\limits_0^Tdt\;\left<m(t)\right>\equiv\oint dk \; \frac{\partial\phi (k)}{\partial k}\pq(k),\label{eq:mkavg} 
\end{align}
where the long-time-averaged probability to find the walker in the reservoir is given by
\begin{equation}
    \label{eq:PAbar}\pq(k) = \lim\limits_{T\rightarrow\infty}\frac{1}{T}\int_0^Tdt\;\pq(k,t).
\end{equation}

The spatially localized initial state [Eq.~\eqref{eq:init_m}] corresponds to a uniform distribution in $k$-space:
\begin{align}
\rho_A(k, 0) &= \left|\psi_A(k,0)\right|^2=\frac{1}{2\pi}\;\forall k,
\label{eq:init_rho}
\end{align}  
which, together with the conservation of $k$, implies the normalization condition $\pp(k)+\pq(k)=1/(2\pi)$. The probability density $\pq(k)=1/(2\pi)-\pp(k)$ controls the behavior of $\left<m\right>$, the central quantity of interest in this paper [see Eq.~\eqref{eq:mkavg}]: if $\pq(k) =1/(2\pi)$ (i.e., $\pp(k)=0$ for all $k$), the walker ends up in the reservoir with certainty in the long-time limit, and the average displacement is simply given by a winding number:
\begin{equation}
\left<m\right> = \frac{\Delta \phi}{2\pi} \equiv\oint \frac{dk}{2\pi} \; \frac{\partial\phi (k)}{\partial k}=\theta(|u|-1),\label{eq:m-winding}
\end{equation} 
where $\theta(x)$ is the Heaviside step function.  Equation \eqref{eq:m-winding} results in a topological transition from $\left<m\right>=0$ to $\left<m\right>=1$ at a critical value $|u|=|v'/v|=1$. This was also the result found for the Markovian quantum walk studied in Ref.~\onlinecite{rudner2009topological}.  
Our goal is now to find the conditions under which the walker decays completely into the reservoir in the long-time limit. 

\section{Mapping to the Weisskopf-Wigner model}\label{sec:equivalent_description}
  In this section, we map the lattice model to the Weisskopf-Wigner model for a two-level atom coupled to radiation modes. This mapping to the Weisskopf-Wigner model gives further insight into the nature of the problem, simplifies the analytical treatment of the master equation, and broadens the possible experimental realizations \cite{krinner2018spontaneous}. 

We start by reformulating  the problem in terms of a single level, $A$, coupled directly with the reservoir. One possible approach is to diagonalize $H_{\bar{A}}(k)$: 
\begin{align}
H(k)&=\epsilon_A\ket{A,k}\bra{A,k}+\sum\limits_{\tilde{j}} \epsilon_{\tilde{j}}\ket{\tilde{j},k}\bra{\tilde{j},k}\nonumber\\&+\sum\limits_{\tilde{j}} \left[\eta_{\tilde{j}}(k)\ket{A,k}\bra{\tilde{j},k}+\mathrm{h.c.}\right], \label{eq:hsw1}
\end{align}
where the levels $\ket{\tilde{j},k}$ are the eigenstates of $H_{\bar{A}}(k)$. If there are $N$ $j$-levels in $C$ there will be $N+1$ levels $\tilde{j}$.

 Another possibility is to apply an approximate Schrieffer-Wolff transformation. For large detuning $\omega$ on site $B$ [Fig.~\ref{fig:lattice}(b)], we can eliminate the coupling to $B$ sites and neglect corrections that are small in $|v(k)|/\omega\ll 1$ and $g_j/\omega\ll 1$ to find a low-energy description of the system (see Appendix \ref{app:SWT}). The resulting Hamiltonian only contains the $A$ sites and the quasi-continuum $C$ with renormalized energies $\tilde{\epsilon}_A$ and $\tilde{\epsilon}_j$, and renormalized coupling 
\begin{equation}
\tilde{\eta}_j(k)= -\frac{v(k)g^*_j}{2}\left(\frac{1}{\omega}+\frac{1}{\omega+\epsilon_A-\epsilon_j}\right).
\end{equation} 
 
 For $v(k)/\omega \ll 1$, we approximate $\tilde{\epsilon}_A(k)\simeq\tilde{\epsilon}_A$, independent of $k$, and set $\tilde{\epsilon}_A=0$.  
 This approximation may break down on a time scale $\tau_A\propto \omega/|v(k)|^2$.
 For this (Schrieffer-Wolff) problem, the long-time limit in our analysis has to be understood as a time long compared to the time scale on which the system reaches the steady state, but short compared to $\tau_A$.  However, in the alternative approach of diagonalizing $H_{\bar{A}}(k)$ and including sites $B$ explicitly in the reservoir [Eq.~\eqref{eq:hsw1}], this approximation would not be necessary.
 
 Finally, the transformed Hamiltonian is given by
\begin{align}
\tilde{H}(k)&=\sum\limits_{j\in C} \tilde{\epsilon}_j\ket{j,k}\bra{j,k}+\sum\limits_{j\in C} \left[\tilde{\eta}_j(k)\ket{A,k}\bra{j,k}+\mathrm{h.c.}\right]. \label{eq:hsw}
\end{align}
Figure \ref{fig:lattice}(b) depicts the level structure described by Eq.~\eqref{eq:hsw}. For each fixed wavenumber $k$, the quantum-walk problem [as described by Eq.~\eqref{eq:hsw}] can be directly mapped to the Weisskopf-Wigner model \cite{scully1999quantum} for an atom decaying into a continuum of radiation modes (the environment): having the walker on site $A$ can be associated with a state where the atom is in the excited state and no photons are in the continuum. We associate the state where the walker is in reservoir state $j\in C$ with an atomic  ground state and one photon in the environment in mode $j$. The formal mapping between these two descriptions is:
\begin{align}
\text{Quantum walk}&\rightarrow \text{Weisskopf-Wigner model}\\
\ket{A,k}&\leftrightarrow \ket{A, k}\otimes \ket{0}\label{eq:models_mapping1}\\
\ket{j,k}&\leftrightarrow \ket{C, k}\otimes \ket{j},\label{eq:models_mapping2}
\end{align}
where $\ket{A,k}=\ket{e}$ labels the atom in the excited state, $\ket{0}$ labels the vacuum state of the environment with 0 photons, $\ket{C,k}=\ket{g}$ labels the atomic ground state, and $\ket{j}$ labels the state of the environment with one photon in state $j$. The quantum walk maps onto a family of Weisskopf-Wigner models, one for each value of $k$, with a $k$-dependent coupling $\tilde{\eta}_j(k)$.

Relative to the quantum walk, there are additional states introduced in the Weisskopf-Wigner model.  These states contain different numbers of excitations: the state $\ket{C, k}\otimes\ket{0}$ with no excitations in the atom or environment, and the states $\ket{A, k}\otimes\ket{j'}$ and $\ket{C, k}\otimes\ket{j'}$, where $j'$ may denote a state with an arbitrary number of excitations in the environment.  The initial state in Eq.~\eqref{eq:init_m} has only one excitation (i.e., the atom is excited and the reservoir is in its vacuum state). Since the Hamiltonian preserves the total number of excitations, the additional states with different numbers of excitations will never become occupied.
Therefore all time-dependent observables evaluated in the Weisskopf-Wigner model with the appropriate initial condition will map identically to observables in the non-Markovian quantum walk. 
The mapping to an enlarged Hilbert space (associated with the Weisskopf-Wigner model) will, however, be useful when evaluating a test of non-Markovianity in Sec.~\ref{sec:nMk}.

Equation \eqref{eq:mkavg} shows that the topological transition, where the average displacement $\left<m\right>$ is quantized, is realized if the walker decays completely into the reservoir, leading to $\rho_{\bar{A}}(k)=1/(2\pi)$. The mapping to the Weisskopf-Wigner model allows us to reformulate the question of a topological transition in terms of a question about a decaying atom. In particular, we ask the question:  Under which conditions does the atom decay to its ground state in the long-time limit?  This question has been addressed previously in the context of the spin-boson model \cite{weiss2012quantum, leggett1987dynamics}. The spin-boson model is a general description of a two-level system (here, $\ket{A, k},\ket{C, k}$) coupled to a bosonic bath. Depending on the spectral density of the bath, the dynamics of the two-level system initialized in $\ket{A, k}$ can either show damped coherent oscillations between states $\ket{A, k}$ and $\ket{C, k}$ with a bounded degree of decay (occupying both $\ket{A, k}$ and $\ket{C, k}$ with a finite probability in the long-time limit), or decay to zero, occupying only $\ket{C, k}$ at long times. In the quantum walk, this second scenario corresponds to the walker leaving sublattice $A$, $\rho_A(k)=0$ ($\rho_{\bar{A}}(k)=1/2\pi$), which in turn leads to the desired topological transition.  The topological transition can therefore be controlled through the reservoir spectral density, as in the case of the spin-boson model \cite{weiss2012quantum, leggett1987dynamics}. 

Note that the above analysis of the Weisskopf-Wigner model assumes an initial zero-temperature reservoir containing no excitations. 
At finite temperature, the reservoir would have a thermal population of photons.
This would lead to absorption that would, in general, lead to a nonvanishing, temperature-dependent stationary excited-state probability for the atom.  
On the other hand, the original formulation of the problem describes a single particle hopping on a lattice, which is out of equilibrium, and thus the notion of temperature does not necessarily apply.

  \section{Long-time dynamics of the walker}\label{sec:ltd}
 The average displacement [Eq.~\eqref{eq:mkavg}] is given in terms of the long-time time averaged probability density $\pq(k)$ [Eq.~\eqref{eq:PAbar}], which in turn can be found from the complementary probability to remain in sublattice $A$: $\pp(k)=1/(2\pi)-\pq(k)$. The Nakajima-Zwanzig generalized master equation \cite{1990_BOOK_Fick,breuer2002theory} can be used to find the time-dependent probability density $\rho_A(k, t)$ without resorting to a Markov approximation. In this section, we work in the language of the Weisskopf-Wigner model, where the state of the walker (atom) is described by the manifold $\left\{\ket{A, k},\ket{C, k}\right\}$ and the state of the environment is independently described by $\left\{\ket{0},\ket{j}\right\}$.  After tracing out the reservoir (environment) degrees of freedom, the Nakajima-Zwanzig generalized master equation gives an equation of motion for the state of the walker (atom) alone.

  The generalized master equation for $\rho_A(k, t)$ in the Born approximation (second order in $\tilde{\eta}_j$) assumes the simple form  (see Appendix \ref{app:GME}):
\begin{equation}
\dot{\rho}_A(k, t) = -i\int\limits^t_0dt'\;\Sigma_A(k, t-t')\rho_A(k, t'),\label{eq:PAdiff}
\end{equation}
where the self-energy $\Sigma_A(k, t)$ is given by:
\begin{align}
\Sigma_A (k, t) &= -2i\sum\limits_j\left|\tilde{\eta}_j(k)\right|^2\cos (\tilde{\epsilon}_jt).\label{eq:selfenergy}
\end{align}
We take the Laplace transform of Eq.~\eqref{eq:PAdiff} [$\tilde{f}(s)=\int_0^\infty dt e^{-st}f(t)$ for any function $f$] and find: 
\begin{equation}
\tilde{\rho}_A(k, s)=\frac{\rho_A(k,0)}{s+i\tilde{\Sigma}_A(k, s)}=\frac{1}{2\pi}\frac{1}{s+i\tilde{\Sigma}_A(k, s)},\label{eq:PAlaplace}
\end{equation}
with 
\begin{align}
\tilde{\Sigma}_A(k, s) &=-2i\int d\epsilon\;J(k,\epsilon)\frac{s}{s^2 +  \epsilon^2},
\label{eq:integralSelfEnergy}
\end{align}
where we have introduced the spectral density $J(k,\epsilon)$:
\begin{equation}\label{eq:spectral_density_discrete}
J(k,\epsilon)\equiv \sum\limits_j\left|\tilde{\eta}_j(k)\right|^2\delta(\epsilon-\tilde{\epsilon}_j).
\end{equation}
To find $\rho_A(k, t)$, we invert Eq.~\eqref{eq:PAlaplace} using the Bromwich inversion integral:
\begin{equation}
\rho_A(k, t)=\frac{1}{2\pi i}\int_{\mathcal{C}}ds\;e^{st}\frac{\rho_A(k, 0)}{s+i\tilde{\Sigma}_A(k, s)},\label{eq:rho_A_t}
\end{equation}
defined on the contour $\mathcal{C}$ running along the vertical line $\mathrm{Re}(s)=\gamma>0$ such that all singularities lie to the left of $\mathcal{C}$. 
The different possible non-analytic features of $\tilde{\rho}_A(k, s)$ give rise to different forms of decay in real time: exponential decay arises from isolated poles at $s=s_j$ with $\mathrm{Re}(s_j)<0$, long-time power-law tails arise from branch-cut integrals, and persistent oscillations arise from poles at $s=s_j$ with $\mathrm{Re}(s_j)=0$.

In terms of the Laplace transform, the long-time time-average of $\rho_A(k, t)$ [$\pp(k)=1/2\pi-\rho_{\bar{A}}(k)$ with $\rho_{\bar{A}}(k)$ defined in Eq.~\eqref{eq:PAbar}] is simply given by the residue at $s=0$:
\begin{eqnarray}
\rho_A(k)&=&\lim \limits_{s\rightarrow 0} s \,\tilde{\rho}_A(k, s)=\frac{1}{2\pi}\frac{1}{1+i\Sigma_A^\prime(k)},\label{eq:longPA}\\
\Sigma_A^\prime(k)&=&\lim_{s\to 0}\frac{\tilde{\Sigma}_A(k,s)}{s}.\label{eq:SigPrime}
\end{eqnarray} 
Contributions from all other terms either decay or oscillate about zero, vanishing under the average given in Eq.~\eqref{eq:PAbar}.

In the continuum limit, the sum in Eq.~\eqref{eq:spectral_density_discrete} is converted to an integral and $J(k,\epsilon)$ becomes a quasi-continuous function.  We distinguish two cases.  When $J(k,\epsilon=\tilde{\epsilon}_A=0)\ne 0$, $\Sigma_A^\prime(k)\to\infty$, and $\rho_A(k)\to 0$ [see Eq.~\eqref{eq:longPA}], i.e., the walker decays completely into the reservoir.  The situation is more complicated when $J(k,\epsilon=\tilde{\epsilon}_A=0)= 0$.  In this case, we assume the low-frequency spectral density is described by a power-law behavior:
\begin{equation}
J(k, \epsilon)=J_\alpha |v(k)|^2\epsilon^\alpha,\quad 0\le\epsilon\le\Delta,\label{eq:spectral_density}
\end{equation}
with spectral exponent $\alpha\ge 0$ and $J_\alpha$ is a dimensionful prefactor. We have introduced a sharp cutoff in the spectral density for frequencies higher than a bandwidth $\Delta>0$.  The spectral exponent characterizes the type of reservoir: sub-Ohmic for $\alpha<1$, Ohmic for $\alpha=1$, and super-Ohmic for $\alpha>1$.

 We insert Eq.~\eqref{eq:spectral_density} into Eq.~\eqref{eq:integralSelfEnergy} and integrate over the finite bandwidth, giving
\begin{equation}
\tilde{\Sigma}_A (k, s) =-i  \frac{\Omega^+_\alpha(k)}{s}\,_2\mathrm{F}_1\left(1, \frac{\alpha+1}{2}; \frac{\alpha+3}{2}; -\frac{\Delta^2}{s^2}\right),
\label{eq:SelfEnergyAlpha}
\end{equation}
 where $_2\mathrm{F}_1\left(a, b; c; z\right)$ is the hypergeometric function and we define
 \begin{equation}
 \Omega^\pm_\alpha(k) = 2J_\alpha\left|v(k)\right|^2 \frac{\Delta^{\alpha\pm 1}}{\alpha\pm 1}.\label{eq:Omega}
 \end{equation}
 Finally, we calculate $\pp(k)$ by inserting Eq.~\eqref{eq:SelfEnergyAlpha} into Eq.~\eqref{eq:longPA}:
 \begin{equation}
\rho_A(k) =\left\{
\begin{array}{ll}
0,&		0\le\alpha\le 1\\\\
\frac{1}{2\pi}\frac{1}{1+ \Omega^-_\alpha(k)},&		\alpha>1.\\
\end{array}\right . \label{eq:PAalfa}
 \end{equation}
The result in Eq.~\eqref{eq:SelfEnergyAlpha} and the subsequent limit in Eq.~\eqref{eq:PAalfa} also arise in the problem of Brownian motion for a quantum particle [Eqs.~(11.1-3) in Ref.~\onlinecite{grabert1988quantum}].

We have found that the walker decays into the continuum if $0\le\alpha\le 1$. Otherwise, there is a residual probability to find the walker on sites $A$ in the long-time limit, given by Eq.~\eqref{eq:PAalfa}. The dissipative dynamics of the spin-boson model has been studied for the spectral density in Eq.~\eqref{eq:spectral_density} \cite{weiss2012quantum, leggett1987dynamics}. At zero temperature, for $\alpha\le 1$ (for an Ohmic or sub-Ohmic spectral density) the spin is localized (the walker ends up in the reservoir), while for $\alpha>1$ (for a super-Ohmic spectral density), the system undergoes persistent coherent oscillations between the two spin states (the walker moves back and forth between the reservoir and the lattice). These results for the spin-boson model are directly transferable to the quantum walk [Eq.~\eqref{eq:PAalfa}].

\begin{figure}[ht]
 \centering
 \includegraphics[width=\columnwidth]{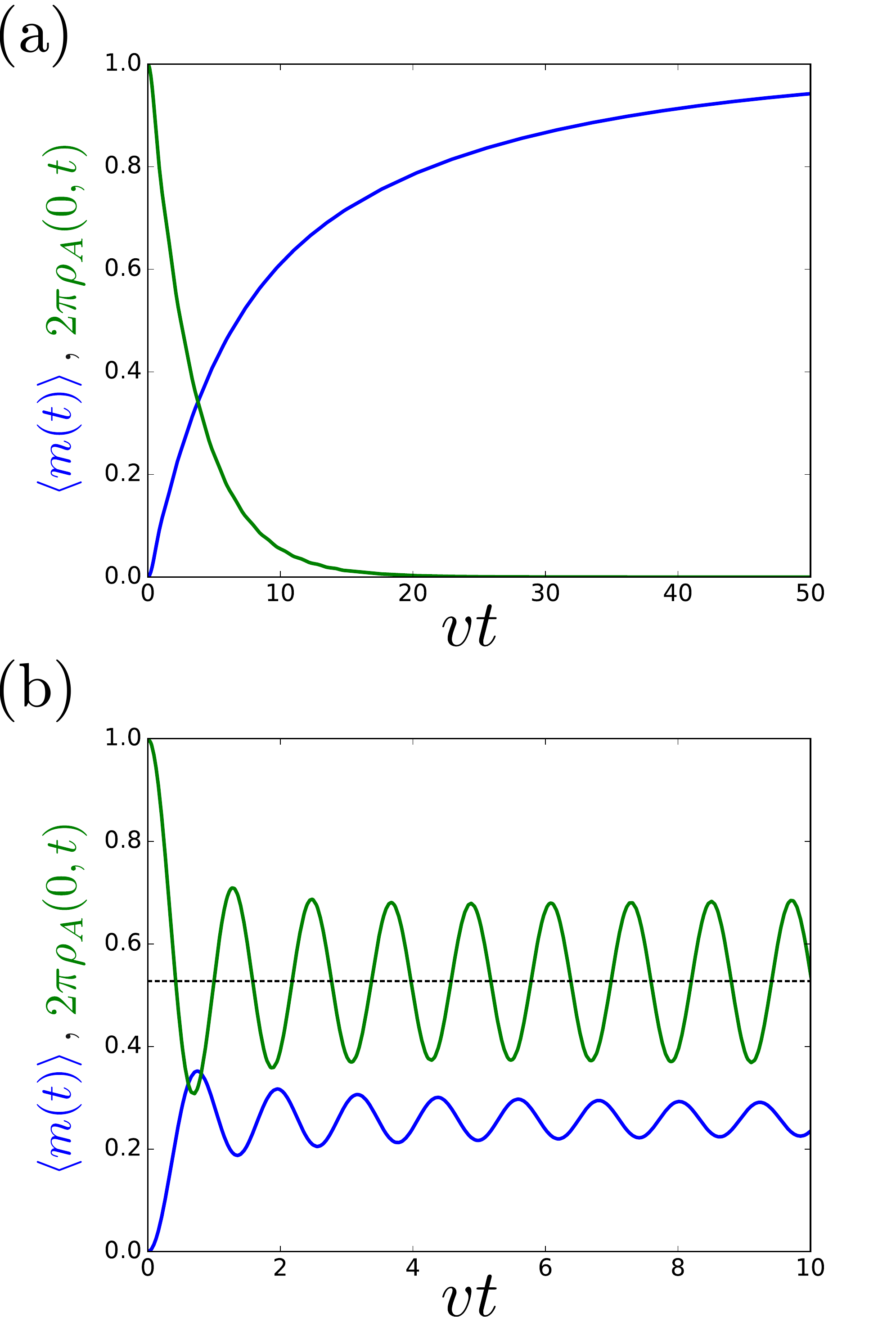}
 \caption{Evolution of the average displacement $\left<m(t)\right>$ (blue lines) and the probability $2\pi\rho_{A}(k=0, t)$ (green lines) as a function of time.  For simplicity, we take $v>0$ and set the dimensionless quantities $J_\alpha v^{\alpha+1}=0.01 $, $\Delta/v=5$, $u=v'/v=2$.  Here, $\alpha = 0$ in (a) and $\alpha=2$ in (b). For $\alpha<1$, the walker decays completely into the reservoir and the average displacement approaches $\left<m(t)\right>\to 1$ at long times for $u=v'/v>1$. For $\alpha>1$, the walker never decays completely into the reservoir and $\rho_{A}(k=0, t)$ oscillates about the long-time time-averaged value [dashed line in (b)] given by Eq.~\eqref{eq:PAalfa}.}\label{fig:probability_time}
 \end{figure}
\begin{figure}[ht]
 \centering
 \includegraphics[width=\columnwidth]{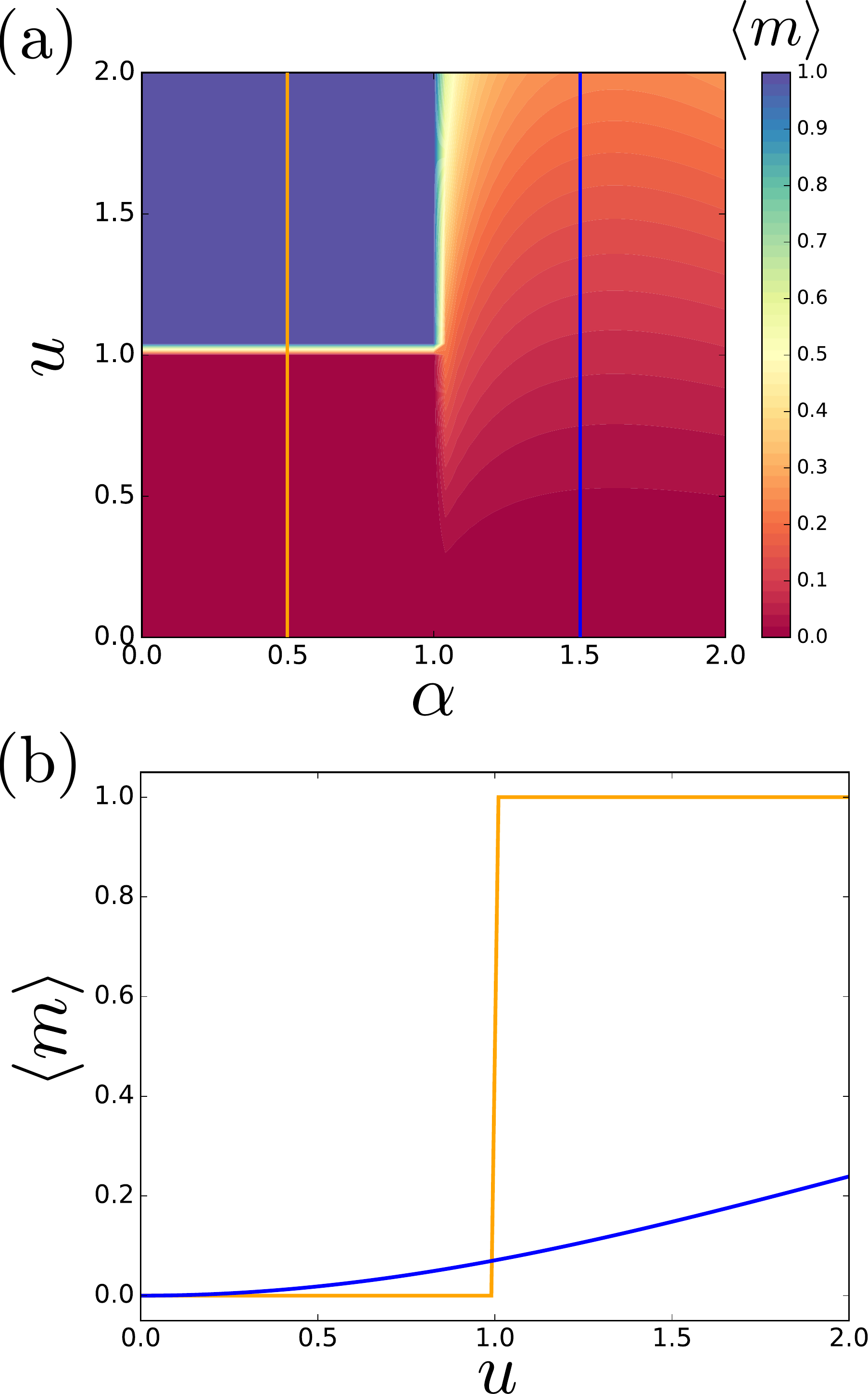}
 \caption{ (a) The phase diagram for the average displacement obtained by integrating Eq.~\eqref{eq:mkavg} using the steady-state probability given by Eq.~\eqref{eq:PAalfa}. For $\alpha<1$ the system displays a robust topological transition at $u=1$, while for $\alpha>1$ there is no transition. 
  (b) Two cuts along the lines of constant $\alpha=0.5$ (orange line) and $\alpha=1.5$ (blue line). Here, $v>0$, $J_\alpha v^{\alpha+1} =0.01$, and $\Delta/v=5$, as in Fig.~\ref{fig:probability_time}.}\label{fig:phase_diagram}
 \end{figure}
In Fig.~\ref{fig:probability_time} we show the probability $2\pi\rho_A(k=0,t)$ obtained by numerical evaluation of the Bromwich integral, and the corresponding time-dependent average displacement $\left<m(t)\right>$. For $\alpha=0$ [Fig.~\ref{fig:probability_time}(a)] the walker completely decays into the reservoir, while for $\alpha=2$ [Fig.~\ref{fig:probability_time}(b)] there is a residual average probability to remain on lattice sites $A$ [the long-time limit, $\rho_A(k)$, as given in Eq.~\eqref{eq:PAalfa}, is indicated with a black dashed line in Fig.~\ref{fig:probability_time}(b)].

The phase diagram in Fig.~\ref{fig:phase_diagram}(a) is the main result of this paper. It shows the average displacement in terms of $u$ and $\alpha$ obtained by numerically integrating Eq.~\eqref{eq:mkavg} using Eq.~\eqref{eq:PAalfa} for $\rho_A(k)=1/2\pi-\rho_{\bar{A}}(k)$. For $0\le \alpha\le 1$, the system has a robust topological transition at $u=1$. The topological phase is not protected for $\alpha>1$, in which case the average displacement is not quantized. In Fig.~\ref{fig:phase_diagram}(b) we show $\left<m\right>$ along two vertical cuts of Fig.~\ref{fig:phase_diagram}(a) (corresponding to $\alpha=0.5$ and $\alpha = 1.5$). This phase diagram extends the result obtained for a Markovian quantum walk in Ref.~\onlinecite{rudner2009topological}.  The extreme Markovian limit in our model corresponds to an infinite-bandwidth flat spectral density ($\alpha=0$, $\Delta\to\infty$).  In this limit, we indeed recover the results of Ref.~\onlinecite{rudner2009topological}, which is to be expected since the two models are equivalent in the Markovian limit.  In contrast with Ref.~\onlinecite{rudner2009topological}, here no Markov approximation was made and we have solved the problem for a generalized spectral function characterized by a general non-negative low-energy spectral exponent $\alpha$. 

  \section{Non-Markovian nature of the quantum walk}\label{sec:nMk}
  
  In this section, we check explicitly that the dynamics of the system is non-Markovian using a method similar to that described in Ref.~\onlinecite{breuer2009measure}, which we summarize here. 

 \begin{figure}[t]
 \centering
 \includegraphics[width=\columnwidth]{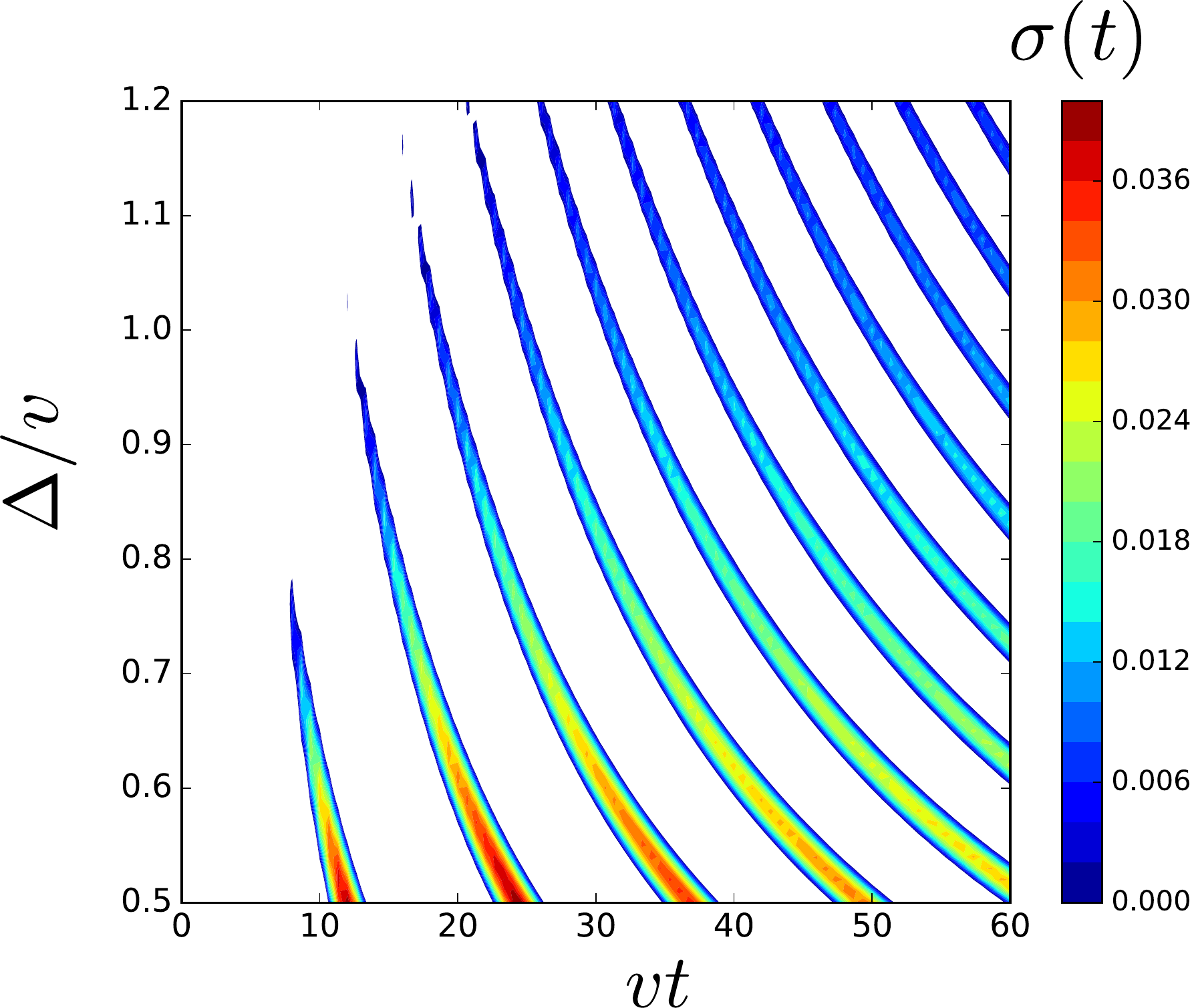}
 \caption{ The non-Markovianity witness given in Eq.~\eqref{eq:non-markovian_condition}, $\sigma(t)\equiv \dot{p}_A(t)$, as a function of the reservoir bandwidth $\Delta$ and time for $\alpha=0.5$. In the blank region, $\sigma\le 0$, while in the shaded regions $\sigma>0$ and the evolution is non-Markovian. Here $v>0$, $J_\alpha v^{\alpha+1}=0.01$ and $u=2$.}\label{fig:non_markovian}
 \end{figure}

The Breuer measure of non-Markovianity \cite{breuer2009measure} for quantum evolution is defined as:
\begin{equation}
\mathcal{N}=\max_{\rho_{1,2}(0)}\int_{\sigma>0}dt\;\sigma\left(t; \uprho_{1,2}(0)\right).\label{eq:measure_NM}
\end{equation}
Here, $\sigma\left(t; \uprho_{1,2}(0)\right)$ is the rate of change of the trace distance between two density matrices $\uprho_{1}(t)$ and $\uprho_{2}(t)$, assumed to evolve according to the same dynamical map:
\begin{equation}
 \sigma\left(t; \uprho_{1,2}(0)\right) = \frac{d}{dt}\mathcal{D}\left[\uprho_1(t), \uprho_2(t)\right],
\end{equation}  
where $\uprho_{1}(0)$ and $\uprho_{2}(0)$ are the initial conditions from which $\uprho_{1}(t)$ and $\uprho_{2}(t)$ evolved.
The trace distance is defined as 
\begin{equation}
\mathcal{D}\left[\uprho_1,\uprho_2\right] = \frac{1}{2}\mathrm{Tr}\left|\uprho_1-\uprho_2\right|,\label{eq:trace_distance}
\end{equation}
where $|A|=\sqrt{A^\dagger A}$. The integral in Eq.~\eqref{eq:measure_NM} is restricted to $\sigma>0$ and the result is maximized over all possible initial conditions $\uprho_{1,2}(0)$.

The trace distance is a monotonic decreasing function of time when the evolution of the density matrices is given by a completely positive and trace preserving (CPT) map. This is the case, for example, for a Markovian process. In these cases, $\sigma$ is never positive:
 \begin{equation}
 \sigma\left(t; \uprho_{1,2}(0)\right) \le 0,
\end{equation}  
and $\mathcal{N}=0$. Thus, to prove that the evolution is non-Markovian, it is sufficient to show that $\sigma$ becomes positive at certain times and for some initial states $\uprho_{1,2}(0)$: $\sigma\left(t; \uprho_{1,2}(0)\right)>0$, yielding $\mathcal{N}>0$. 

We work with the reduced density matrix obtained by marginalizing over all wavevectors $k$:
\begin{align}\label{eq:rho12}
\uprho_{1,2}(t)&=\sum\limits_{l,l'}\ket{l}\bra{l'}\oint dk\;\mathrm{Tr}_\mathrm{E}\left\{\bra{l,k}\rho_{1,2}(t)\ket{l',k}\right\},
\end{align}
where $\mathrm{Tr_E}$ is the partial trace over the environment degrees of freedom describing the photon state for the Weisskopf-Wigner problem, $\mathrm{Tr_E}\left \{\cdot\right\}\equiv \bra{0}\cdot\ket{0}+\sum\limits_j\bra{j}\cdot\ket{j}$. Here, $l,l' \in \{A,C\}$ label quantum numbers for a fictitious two-dimensional auxiliary space that makes it convenient to calculate the total probability $p_A(t)$ for the walker to be found in the sublattice $A$ at time $t$. 

The state $\rho_{1,2}(t)$ in Eq.~\eqref{eq:rho12} is the full density matrix in the enlarged Hilbert space of the Weisskopf-Wigner problem at time $t$, conditioned on the initial states:
\begin{eqnarray}
\rho_1(0) & = & \ket{A, m=0}\bra{A, m=0}\otimes\ket{0}\bra{0},\label{eq:rho1_0}\\
\rho_2(0) & = & \ket{C, m=0}\bra{C, m=0}\otimes\ket{0}\bra{0}.\label{eq:rho2_0}
\end{eqnarray}
The initial condition $\rho_1(0)$ describes a uniform coherent superposition of Weisskopf-Wigner problems with different $k$ corresponding to an atomic excited state $A$ with no photons in the environment ($\ket{0}$). This is equivalent to having the walker initialized to sublattice $A$, localized at lattice site $m=0$ [
Eq.~\eqref{eq:init_m}].  We compare evolution with this initial condition to the trivial evolution found with the initial condition $\rho_2(0)$.  The state $\rho_2(0)$ corresponds instead to having the atom in the ground state and no photons in the environment in the Weisskopf-Wigner model.  This is equivalent to having no walker. Since $\rho_1(0)$ and $\rho_2(0)$ in Eqs.~\eqref{eq:rho1_0}, \eqref{eq:rho2_0} describe orthogonal pure states, this choice of two initial states maximizes the trace distance, Eq.~\eqref{eq:trace_distance}, at $t=0$.

The time evolution of $\uprho_2(t)$ is trivial, $\uprho_2(t)=\uprho_2(0)$, since the initial condition describes a state with no excitations (no walker).  For the initial condition $\uprho_1(0)$, we find the future evolution of $\uprho_1(t)$ is given by (see Appendix \ref{app:GME}):
\begin{align}
\uprho_1(t) &=p_A(t)\ket{A}\bra{A}+\left[1-p_A(t)\right]\ket{C}\bra{C},\\
p_A(t)&=\oint dk\;\rho_A(k,t),
\end{align}
where the total probability for the walker to be on sublattice site $A$, $p_A(t)$, is simply found by integrating the probability density $\rho_A(k,t)$ over $k$.

From Eq.~\eqref{eq:trace_distance} we find that the trace distance can be expressed directly in terms of $p_A(t)$:
\begin{align}
\mathcal{D}\left[\uprho_1(t),\uprho_2(t)\right]\equiv p_A(t).\label{eq:Dtrace}
\end{align}
It follows that the dynamics of the system is non-Markovian ($\mathcal{N}>0$) if:
\begin{equation}
\sigma(t)=\dot{p}_A(t)=\oint dk\;\dot{\rho}_A(k,t)>0
\label{eq:non-markovian_condition}
\end{equation} 
for some time $t$.

In Fig.~\ref{fig:non_markovian} we plot the non-Markovianity witness, $\sigma(t)=\dot{p}_A(t)$, found from numerical integration of Eq.~\eqref{eq:rho_A_t} with the initial condition given in Eq.~\eqref{eq:init_rho}. 
A positive value of the witness, $\sigma(t)=\dot{p}_A(t)>0$, indicates non-Markovian dynamics (the walker returns to sublattice $A$ from the reservoir). 
The result in Fig.~\ref{fig:non_markovian} is shown as a function of the reservoir bandwidth $\Delta$ and time $t$: shaded regions clearly indicate non-Markovian evolution even in the topologically protected part of the phase diagram in Fig.~\ref{fig:phase_diagram}(a) ($\alpha\le 1$). 
This demonstrates that Markovianity is not essential for the topological transition. Furthermore, as one would intuitively expect, the evolution becomes more Markovian for larger bandwidth $\Delta$.  In particular, for any fixed time $t$, Fig.~\ref{fig:non_markovian} shows $\sigma\le 0$  for sufficiently large bandwidth $\Delta$.

  \section{Conclusions and outlook}\label{sec:con}
We have extended the analysis presented in Ref.~\onlinecite{rudner2009topological} for Markovian non-Hermitian quantum walks by introducing an additional set of levels in each unit cell describing a reservoir that the walker may coherently enter and leave. 
The robust topological transition in the average walker displacement $\left<m\right>$, previously found for the Markovian model, persists in the non-Markovian regime, under appropriate conditions on the reservoir spectral density.  The result is summarized in the phase diagram presented in Fig.~\ref{fig:phase_diagram}.  In particular, this phase diagram shows that for a sub-Ohmic or Ohmic spectral density [$J(\epsilon)\propto \epsilon^\alpha$ with spectral exponent $\alpha\le 1$], the walker decays into the reservoir in the long-time limit, and there is a robust topological transition; for a super-Ohmic spectral density ($\alpha>1$), there is no transition.  This difference in walker dynamics is directly related to the well-known localization transition in the spin-boson model \cite{leggett1987dynamics}.

To confirm that the walker dynamics is non-Markovian, we have numerically evaluated a non-Markovianity witness based on the Breuer measure \cite{breuer2009measure} of non-Markovianity. In this problem, the non-Markovianity witness is positive whenever the walker re-enters the lattice from the continuum.  The result is shown in Fig.~\ref{fig:non_markovian}, where the dynamics is clearly non-Markovian even in a parameter regime with a robust topological transition.

In addition to a direct simulation of the quantum walk [Fig.~\ref{fig:lattice}(a)] through an array of, e.g., quantum dots, it may also be possible to experimentally realize the equivalent level structure shown in Fig.~\ref{fig:lattice}(b) with an alternative system that may be easier to engineer and control.  For example, the starting Hamiltonian presented in Eqs.~\eqref{eq:HAk}-(\ref{eq:Vk}) is analogous to that used routinely in a STIRAP setup \cite{vitanov2017stimulated}.  Alternatively, we have shown that the problem of the walker can be mapped to the Weisskopf-Wigner problem of atomic decay with a structured environment. In a recent work  (Ref.~\onlinecite{krinner2018spontaneous}), Krinner et al.~have simulated Weisskopf-Wigner dynamics with an engineered spectral density. Their implementation used ultracold atoms in an optical lattice, and they have observed non-Markovian dynamics in the system.
Executing such an experiment with a range of parameters $v(k)$ for $k\in(0,\pi)$ would allow one to reconstruct the functional form of $\rho_{\bar{A}}(k)$ throughout the Brillouin zone. This would allow for an evaluation of the average walker displacement $\left<m\right>$ [see Eq.~\eqref{eq:mkavg}]. If the experiment were reproduced for a range of spectral exponents $\alpha$ and ratios $u=v'/v$, the phase diagram shown in Fig.~\ref{fig:phase_diagram} could potentially be confirmed experimentally.

More broadly, this work extends the notion of non-equilibrium topological phenomena to non-Markovian systems, and stimulates the question: what new robust non-equilibrium phenomena may be found in non-Markovian, open quantum systems?

\begin{acknowledgments}
WAC and AR acknowledge funding from the Natural Sciences and Engineering Research Council (NSERC) and the Fonds de Recherche du Québec–Nature et technologies (FRQNT). MR is grateful for the support of the European Research Council (ERC) under the European Union Horizon 2020 Research and Innovation Programme, Grant Agreement No. 678862, and the Villum Foundation.
\end{acknowledgments}

\appendix

\section{Average Displacement}\label{app:avgD}

We start from the definition of the average displacement given in Eq.~\eqref{eq:m_m}. We apply the Fourier transform defined above Eq.~\eqref{eq:HAk} and we find
\begin{align}
\left<m(t)\right>&=\oint \frac{dk}{\sqrt{2\pi}} \oint\frac{dk'}{\sqrt{2\pi}}\sum\limits_{l\notin A}\psi_{l}^*(k',t)\psi_{l}(k,t)\nonumber\\
&\qquad\qquad\times\sum\limits_m\left[-i\partial_k e^{im(k-k')}\right]\\
&=\oint \frac{dk}{\sqrt{2\pi}}\oint\frac{dk'}{\sqrt{2\pi}}\sum\limits_{l\notin A}\psi_{l}^*(k',t)\left[i\partial_k\psi_{l}(k,t)\right]\nonumber\\
&\qquad\qquad\times\sum\limits_m e^{im(k-k')}\\
&=i\oint dk\sum\limits_{l\notin A}\psi_{l}^*(k,t)\partial_k\psi_{l}(k,t),
\end{align}
where we integrated by parts.
We now rotate away the phase $\phi(k)$ associated with the Fourier transform of the coupling $v(k)=|v(k)|e^{i\phi(k)}$:
\begin{equation}
H(k)\rightarrow \bar{H}(k)=e^{i\Pi_{\bar{A}}\phi(k)}H(k)e^{-i\Pi_{\bar{A}}\phi(k)},\label{eq:rotated_H}
\end{equation}
where $\Pi_{\bar{A}}$ is the projector onto the $\bar{A}$ subspace:
\begin{equation}
\Pi_{\bar{A}}=\mathbf{1}-\ket{A}\bra{A}.
\end{equation}
The states transform as
\begin{equation}
\ket{\psi(t)}\rightarrow \ket{\bar{\psi}(t)}=e^{i\Pi_{\bar{A}}\phi(k)}\ket{\psi(t)}.
\end{equation}
Then, in this rotated frame, we find
 \begin{align}
\left<m(t)\right>&=i\oint dk\sum\limits_{l\notin A} e^{i\phi(k)}\bar{\psi}_{l}^*(k,t)\nonumber\\
&\times\partial_k\left[e^{-i\phi(k)}\bar{\psi}_{l}(k,t)\right]\\
&=i\oint dk\sum\limits_{l\notin A} \bar{\psi}_{l}^*(k,t)\partial_k\bar{\psi}_{l}(k,t)\nonumber\\
&+\oint dk\sum\limits_{l\notin A}\left|\bar{\psi}_{l}(k,t)\right|^2\partial_k\phi(k).\label{eq:m_app}
\end{align}

The transformed Hamiltonian $\bar{H}(k)$ is symmetric under parity, $\bar{H}(k)=\bar{H}(-k)$, and the evolution under $\bar{H}(k)$ therefore preserves parity. Then, since the initial state is even under parity, $\bar{\psi}_l(k,0)=\bar{\psi}_l(-k,0)$, the state at any future time will also be even: $\bar{\psi}_l(k,t)=\bar{\psi}_l(-k,t)$, and the first term in Eq.~\eqref{eq:m_app} is zero because the integrand is odd. Finally, using Eq.~\eqref{eq:escapeProb} we find the result given in Eq.~\eqref{eq:mk} of the main text:
\begin{align}
\left<m(t)\right>=\oint dk \; \frac{\partial\phi(k)}{\partial k}\pq(k,t).
\end{align}
 \section{Schrieffer-Wolff approximation}\label{app:SWT}
We split the Hamiltonian $H(k)$ into a diagonal term $H_0(k)$ and terms coupling $B$ to $A$ sites $V(k)$ [Eq.~\eqref{eq:Vk}] and $B$ and $\{j\}$ sites $V_g(k)$: 
\begin{align}
H(k)&=H_0(k)+V(k)+V_g(k)\\
H_0(k) &= \epsilon_A \ket{A,k}\bra{A,k}+ (\epsilon_A+\omega) \ket{B,k}\bra{B,k}\nonumber\\
&+  \sum_{j\in C}\epsilon_j\ket{j,k}\bra{j,k}\\
V_g(k) &=  \sum_{j\in C}  g_j\ket{j,k}\bra{B,k} + \mathrm{h.c.}.
\end{align}

We rotate the Hamiltonian by means of an anti-Hermitian generator $S$
\begin{align}
H(k)&\rightarrow  e^SH(k)e^{-S}\\
&= H(k) + [S, H(k)] + \frac{1}{2}[S, [S, H(k)]] + \cdots.
\end{align}
 We eliminate the coupling to $B$ sites at first order by imposing $[S, H_0(k)] = -\left[V(k) + V_g(k)\right]$. Explicitly, $S$ is given by:
 \begin{align}
     S&=-\frac{v}{\omega}\ket{A, k}\bra{B, k} -\mathrm{h.c.}\nonumber\\
     &-\sum\limits_{j}\frac{g_j}{\epsilon_A+\omega-\epsilon_j}\ket{j,k}\bra{B,k}-\mathrm{h.c.}.
 \end{align} 
 We keep only terms up to second order in $V(k)$ and $V_g(k)$ and project out site $B$. 
 The resulting Hamiltonian $H_\mathrm{SW}(k)$ has renormalized energies on sites $A$ and $\{j\}$, as well as a direct coupling between site $A$ and levels $\{j\}$.  We find:
\begin{align}
H_\mathrm{SW}(k) &= H_0(k)+\frac{1}{2}[S, V(k)] + \frac{1}{2}[S, V_g(k)],\\
&=\tilde{\epsilon}_A(k)\ket{A,k}\bra{A,k}+\sum\limits_j \tilde{\epsilon}_j\ket{j,k}\bra{j,k}\nonumber\\&+\sum\limits_j \tilde{\eta}_j(k)\ket{A,k}\bra{j,k}+h.c.\nonumber\\
&+\sum\limits_{j\neq j'}t_{jj'}(k)\ket{j,k}\bra{j',k}+h.c.,\label{eq:lowenergyH}\\
\tilde{\epsilon}_A(k)&=\epsilon_A - \frac{|v(k)|^2}{\omega},\label{eq:epsilontildeA}\\
\tilde{\epsilon}_j&=\epsilon_j- \frac{1}{2}\frac{|g_j|^2}{\epsilon_A+\omega-\epsilon_j},\\
\tilde{\eta}_j(k)&=-\frac{v(k)g^*_j}{2}\left(\frac{1}{\omega}+\frac{1}{\epsilon_A+\omega-\epsilon_j}\right),\\
t_{jj'}&=-\frac{1}{2}\frac{g_jg^*_{j'}}{\epsilon_A+\omega-\epsilon_j}.
\end{align}

The coupling $t_{jj'}$ between levels in the reservoir will lead to a modification of the spectral density in general.  When a particular spectral density can be designed for the bare levels $\{j\}$, it may be advantageous to work in the limit of weak coupling $g_j$ and large $\omega$, so that these corrections can be neglected.  Alternatively, the reservoir could be re-diagonalized accounting for these corrections, and it may be possible to account for the associated change to the spectral density.

\section{Generalized master equation}\label{app:GME}
We want to write a master equation for the probability density $\rho_{A}(k, t)$ of finding the walker in sublattice $A$ (having the atom in the excited state, $A$, in the Weisskopf-Wigner model):
\begin{align}
\rho_{A}(k, t) &= \mathrm{Tr}\left\{ \ket{A,k}\bra{A,k}\rho(t) \right\},\label{eq:pa_general}
\end{align}
where here we take $\rho(t)$ to be the full density matrix in the enlarged (tensor product) Hilbert space of the Weisskopf-Wigner model. In this description, the Hilbert space of the walker (atom) is spanned the states $\ket{A, k}$ and $\ket{C, k}$ and the Hilbert space of the reservoir is spanned by the states $\ket{0}$ and $\{\ket{j}\}$ [Eqs.~\eqref{eq:models_mapping1} and \eqref{eq:models_mapping2}].

We assume an initial tensor product state:
\begin{equation}
    \rho(0)=\rho_\mathrm{S}(0)\otimes \rho_\mathrm{E}(0),\label{eq:initialApp}
\end{equation}
where $\rho_\mathrm{S}(0)$ denotes the initial density matrix of the system (walker or atom) and $\rho_\mathrm{E}(0)$ is the initial density matrix of the environment, which we take to be
\begin{equation}
\rho_\mathrm{E}(0) = \ket{0}\bra{0},\label{eq:init_env}
\end{equation}
where the state $\ket{0}$ indicates the vacuum state of the environment with no photons.

We project out the degrees of freedom of the environment to obtain a master equation for the density matrix $\rho_\mathrm{S}(t)$ of the system. To do this, we use the projection superoperator $\mathcal{P}$, defined as
\begin{equation}
\mathcal{P}\cdot = \mathrm{Tr_E}\left \{\cdot \right\}\otimes \rho_\mathrm{E}(0),
\end{equation}
where $\mathrm{Tr_E}$ is the partial trace over the environment degrees of freedom, $\mathrm{Tr_\mathrm{E}}\left \{\cdot\right\}\equiv \bra{0}\cdot\ket{0}+\sum\limits_j\bra{j}\cdot\ket{j}$.
We also define the Liouvillian superoperators $\mathcal{L}_\mathrm{S}$ and $\mathcal{L}_V$:
\begin{align}
    \mathcal{L}_\mathrm{S} \cdot &= [H_\mathrm{S},\cdot]\\
    \mathcal{L}_V(t) \cdot &= [V(t),\cdot],
    \end{align}
    where the system and coupling Hamiltonians in the Weisskopf-Wigner Hilbert space are defined by:
    \begin{align}
    H_\mathrm{S} &= \oint dk\;\ket{C,k}\bra{C,k}\otimes\sum\limits_j\tilde{\epsilon}_j\ket{j}\bra{j},\\
    V(t)&=\sum\limits_j \oint dk\;\tilde{\eta}_{j}(k)e^{-i\tilde{\epsilon}_jt}\ket{A,k}\bra{C,k}\otimes\ket{0}\bra{j}+\mathrm{h.c.}. \label{eq:Vnew}
\end{align}

For the initial condition given in Eq.~\eqref{eq:initialApp}, $\mathcal{P}$ has the following properties:
\begin{align}
\mathcal{P}\rho(t) &= \mathrm{Tr_E}\left \{\rho(t) \right\}\otimes \rho_\mathrm{E}(0)\nonumber\\
&=\rho_\mathrm{S}(t)\otimes \rho_\mathrm{E}(0)\\
\mathcal{P}\rho(0) &= \rho(0)\\
\mathrm{Tr}\left\{ \ket{l, k}\bra{l, k}\mathcal{P}\rho(t) \right\}&=\mathrm{Tr}\left\{ \ket{l, k}\bra{l, k}\rho(t) \right\}\nonumber\\
&=\rho_{l}(k, t),
\end{align} 
where $l\in\{A,C\}$.

Using these properties, and noting that $\mathcal{P}\mathcal{L}_V\mathcal{P}=0$, the generalized master equation for $\mathcal{P}\rho( t)$ in the interaction picture is given by \cite{breuer2002theory}:
\begin{equation}
\frac{d}{dt}\mathcal{P}\tilde{\rho}(t) = -i\int\limits^t_0dt'\;\Sigma(t,t')\mathcal{P}\tilde{\rho}(t').
\end{equation}
Here, $\tilde{\rho}(t)$ is the density matrix in the interaction picture with respect to the system Hamiltonian:
\begin{equation}
    \tilde{\rho}(t)=e^{i\mathcal{L}_\mathrm{S}t}\rho(t).
\end{equation}
In the Born approximation, the self-energy $\Sigma(t,t')$ is given by:
\begin{align}
\Sigma(t,t')&=-i\mathcal{P}\mathcal{L}_V(t)\mathcal{L}_V(t')\mathcal{P}. 
\label{eq:selfenergyA}
\end{align}

Expanding Eq.~\eqref{eq:selfenergyA} in terms of commutators, we find:
\begin{equation}
\dot{\rho}_\mathrm{S}(t) = -\int\limits^t_0dt'\;\mathrm{Tr_E}\left \{[V(t),[V(t'),\rho_\mathrm{S}(t') \otimes\rho_\mathrm{E}(0)]]\right\}.\label{eq:master_rhoS}
\end{equation}

Evaluating the commutators and using 
\begin{equation}
\rho_{A}(k, t)= \bra{A,k}\rho_{S}(t)\ket{A,k},
\end{equation}
we find Eqs.~\eqref{eq:PAdiff} and \eqref{eq:selfenergy} in the main text.
\bibliography{TopolTransNMQmWalk}
\clearpage
\end{document}